\documentclass[prd, twocolumn, nofootinbib, floatfix]{revtex4-1}

\usepackage{amsmath}
\usepackage{graphicx}
\usepackage{dcolumn}
\usepackage{bm}
\usepackage{epsfig}
\usepackage{amssymb,latexsym,mathrsfs}
\usepackage{graphicx}
\usepackage{color}
\usepackage{hyperref}
\usepackage{float}
\usepackage{diagbox}  
\usepackage{cancel} 
\usepackage[dvipsnames]{xcolor}

\hypersetup{
    colorlinks=true,
    linkcolor=red,
    citecolor=blue,
} 

\newcommand{\be}{\begin{equation}}
\newcommand{\ee}{\end{equation}}
\newcommand{\bs}{\begin{split}} 
\newcommand{\bea}{\begin{eqnarray}}
\newcommand{\eea}{\end{eqnarray}}

\newcommand{\kp}{\kappa}

\newcommand{\eps}{\epsilon}
 
\newcommand{\bsq}{|\beta_{\omega\omega'}|^2}
\newcommand{\bww}{\beta_{\omega\omega'}}

\begin{document}

\title{M\"obius Mirrors} 
\author{Michael R.R. Good${}^{1,2}$}
\author{Eric V.\ Linder${}^{2,3}$} 
\affiliation{${}^1$Physics Department, Nazarbayev University, Nur-Sultan, 010000
Kazakhstan\\
${}^2$Energetic Cosmos Laboratory, Nazarbayev University, Nur-Sultan, 010000
Kazakhstan\\ 
${}^3$Berkeley Center for Cosmological Physics \& Berkeley Lab, 
University of California, Berkeley, CA 94720, USA} 

\begin{abstract} 
An accelerating boundary (mirror) acts as a horizon and black hole 
analog, radiating energy with some particle spectrum. We demonstrate 
that a M\"obius transformation on the null coordinate advanced time mirror trajectory 
uniquely keeps invariant not 
only the energy flux but the particle spectrum.  
We clarify how the geometric entanglement entropy is also invariant. The transform allows generation 
of families of dynamically distinct trajectories, including 
$\mathcal{PT}$-symmetric ones, mapping from the eternally thermal 
mirror to the de Sitter horizon, and different boundary motions  corresponding to Kerr or Schwarzschild black holes. 
\end{abstract} 

\date{\today} 

\maketitle


\section{Introduction} 

Black holes represent a fascinating intersection of spacetime 
and information, probing the nature of each, and their 
relation. The relation is intimately tied with the horizon, and 
one can study other systems with horizons to elucidate the 
connections.  Using the moving mirror model \cite{DeWitt:1975ys,Davies:1976hi,Davies:1977yv}, the accelerating boundary correspondence (e.g.\ Schwarzschild \cite{Good:2016oey}, Reissner-Nordstr\"om \cite{good2020particle}, Kerr \cite{Good:2020fjz}) maps 
hot black holes to particular moving mirror trajectories with horizons.  While a close connection between the particle and energy creation by moving mirrors \cite{walker1985particle,unruhwald1982} and black holes has been known for sometime, complete investigation of these exact analogs are ongoing. Quasi-thermal solutions typify geometric end-states  like black hole remnants \cite{Chen:2014jwq} (asymptotic constant-velocity mirrors \cite{Good:2016atu,Good:2018ell,Good:2018zmx,Good:2015nja,Good:2016yht}). Complete black hole evaporation models are characterized by asymptotic zero-velocity mirrors \cite{Walker_1982, Good:2019tnf,GoodMPLA,Good:2017kjr,B,Good:2017ddq,Good:2018aer}. 

While entanglement entropy \cite{Holzhey:1994we}, and hence information, is tied 
directly to the mirror trajectory, the distant observer only 
detects energy flux and particle production. We investigate 
the connection between these by considering special 
transformations of the mirror trajectory such that the energy 
flux remains invariant. These are M\"obius transformations arising from the Schwarzian operator in the quantum stress 
energy tensor, and correspond to 
the $SL(2,\mathbb{R})$ group symmetry. 

Using this, we can explore the relation between mirrors, and 
hence spacetimes, with identical flux, such as thermal emission 
from de Sitter space \cite{Good:2020byh} and the Carlitz-Willey \cite{carlitz1987reflections} mirror, 
investigate cases with merely asymptotically identical flux, 
and probe the zero energy, but nonzero particle, production of 
uniformly accelerating motion, whose asymptotic dynamics corresponds to extremal black holes \cite{Liberati:2000sq,good2020extreme,Good:2020fjz,Rothman:2000mm,Foo:2020bmv}.

In Sec.~\ref{sec:mob} we describe the M\"obius transform and 
its effects, applying it to eternally thermal flux and Planckian  
particle production in Sec.~\ref{sec:eternal}. We consider the 
de Sitter spacetime in particular in Sec.~\ref{sec:dS},  
Schwarzschild and Kerr black holes in Sec.~\ref{sec:schw}, and 
the uniform acceleration case in Sec.~\ref{sec:uniform}, concluding 
in Sec.~\ref{sec:concl}.

 \section{M\"obius Transformations} \label{sec:mob} 
 
 The quantum stress tensor for an accelerating boundary 
 correspondence (ABC) indicates an energy flux $\mathcal{F}$ produced from 
 the boundary (horizon) to an observer at infinity,  
 \be 
 -24\pi \mathcal{F}(u) = \{p,u\} \equiv \frac{p'''}{p'} - \frac{3}{2}\left(\frac{p''}{p'}\right)^2\,, 
 \ee 
 where $p(u)$ is the trajectory, i.e.\ mirror position $v$ as 
 a function of $u$, where $u=t-x$ and $v=t+x$ are the null 
 coordinates, also called the retarded and advanced times. It can be derived via point-splitting \cite{Davies:1976hi}, or via the Schwinger term in the Virasoro algebra \cite{Reuter:1988nt}. 
 We use natural units, e.g.\ $\hbar = k_B = 1$,  throughout. 
 
 The notation $\{p,u\}$ denotes the Schwarzian derivative, 
 and this signals that an ABC has an underlying symmetry in the 
 form of the M\"obius transformations of $SL(2,\mathbb{R})$,
\be 
p(u) \to \frac{ a p(u) + b }{c p(u) + d}, \quad ad-bc = 1\,.  
\ee
(Since $p$ nominally has spacetime dimension one, we can 
make it dimensionless by interpreting it as $\kp p$, with 
$\kp$ a normalizing factor.) 

Thus, a trajectory 
\be 
P(u)=\frac{ a p(u) + b }{c p(u) + d} \equiv M p(u)\, 
\ee 
with $ad-bc=1$ has the same energy flux as the original $p(u)$. 
Later, we will show that it has invariant particle 
flux seen by an observer as well. 

We can divide the transforms into two cases, when $c=0$ and when 
$c\ne0$. When $c=0$ then $P=(a/d)p+b/d$. The $b/d$ piece is a 
shift in $p$, the translation part of the M\"obius transform. 
This does not contribute to $P'$ and hence not to the energy 
flux, and since $P$ enters 
the Bogolyubov beta coefficient 
$\beta_{\omega\omega'}$ 
as $e^{-i\omega'P}$ then a constant addition in $P$ is a pure phase 
and will cancel in $\bsq$. 
Nor does the translation change the rapidity $\eta(u)=(1/2)\ln p'(u)$ 
and so it is not of much interest. It can, however, alter the 
position of the mirror at infinity, e.g.\ whether the mirror 
starts at null coordinate $v=0$ or some finite value. 

Thus the $c=0$ case (hence $d=1/a)$ is 
\be 
P=a^2p\qquad (c=0)\,. \label{eq:pc0}
\ee 
This represent the dilatation part of the M\"obius transform. 
Note that since $\eta(u)=(1/2)\ln p'$ then a constant 
multiplicative factor for $P$ is just an additive 
shift of $\eta\to\eta+\ln a$.  This does have physical 
consequences, as an additive factor to the entropy flux, 
and a multiplicative factor to the mirror acceleration. 

When $c\ne0$ then we can multiply the numerator and denominator in 
$P$ by $c$ to get 
\be 
P=\frac{acp+ad-ad+bc}{c(cp+d)}=\frac{a}{c}-\frac{1}{c(cp+d)}\,, 
\ee 
where we have used $ad-bc=1$. Again ignoring the constant term, 
we can write ($s\equiv cd$) 
\be 
P=\frac{-1}{c^2p+s} \qquad (c\ne0)\,. \label{eq:pc} 
\ee 
Now $\eta(P)$ is not simply related to $\eta(p)$, and 
the entropy and acceleration are likewise different 
for the two trajectories, though the energy flux is the same. 
When $d=0$ ($s=0$) then the M\"obius transform is an inversion of $p$, with 
a minus sign.

Equations~\eqref{eq:pc0} and \eqref{eq:pc} are the two transforms 
exhibiting the symmetry of the ABC due to the Schwarzian, hence 
leaving the energy flux invariant. We are also interested in 
whether this carries over to the particle creation and its 
spectrum. Recall that the particle creation per mode per mode is 
$N_{\omega\omega'}=\bsq$, where $\omega$ is the outgoing and 
$\omega'$ the ingoing frequency, and the particle spectrum 
seen by a distant observer is 
$N(\omega)=\int_0^\infty d\omega'\,\bsq$. In particular, for 
the thermal case $N(\omega)$ has a Planckian spectrum of particles. 
The Bogolyubov beta coefficient follows   
\be 
\bww = \frac{-1}{2\pi}\sqrt{\frac{\omega}{\omega'}}\int_{u_{\textrm{min}}}^{u_{\textrm{max}}}du\, e^{-i\omega u-i\omega'p(u)}\ .  \label{eq:bww} 
\ee 
While we cannot evaluate this for $P(u)$ coming from arbitrary 
$p(u)$, we examine several physically important cases in the following 
sections.

\section{Eternal Thermal Mirror} \label{sec:eternal} 

We begin with the classic case of constant thermal energy 
emission, having particles distributed in a Planck distribution as 
discovered by Carlitz-Willey  \cite{carlitz1987reflections} (see \cite{Good:2017ddq,Good:2012cp,B} for its trajectory in spacetime coordinates and further detail). 
This corresponds to a particular nonuniformly accelerating mirror, 
given by 
\be 
p(u) = \frac{-1}{\kp}\,e^{-\kp u}\ . \label{eq:CW} 
\ee 

\begin{figure}[h]
\centering 
\includegraphics[width=3.0in]{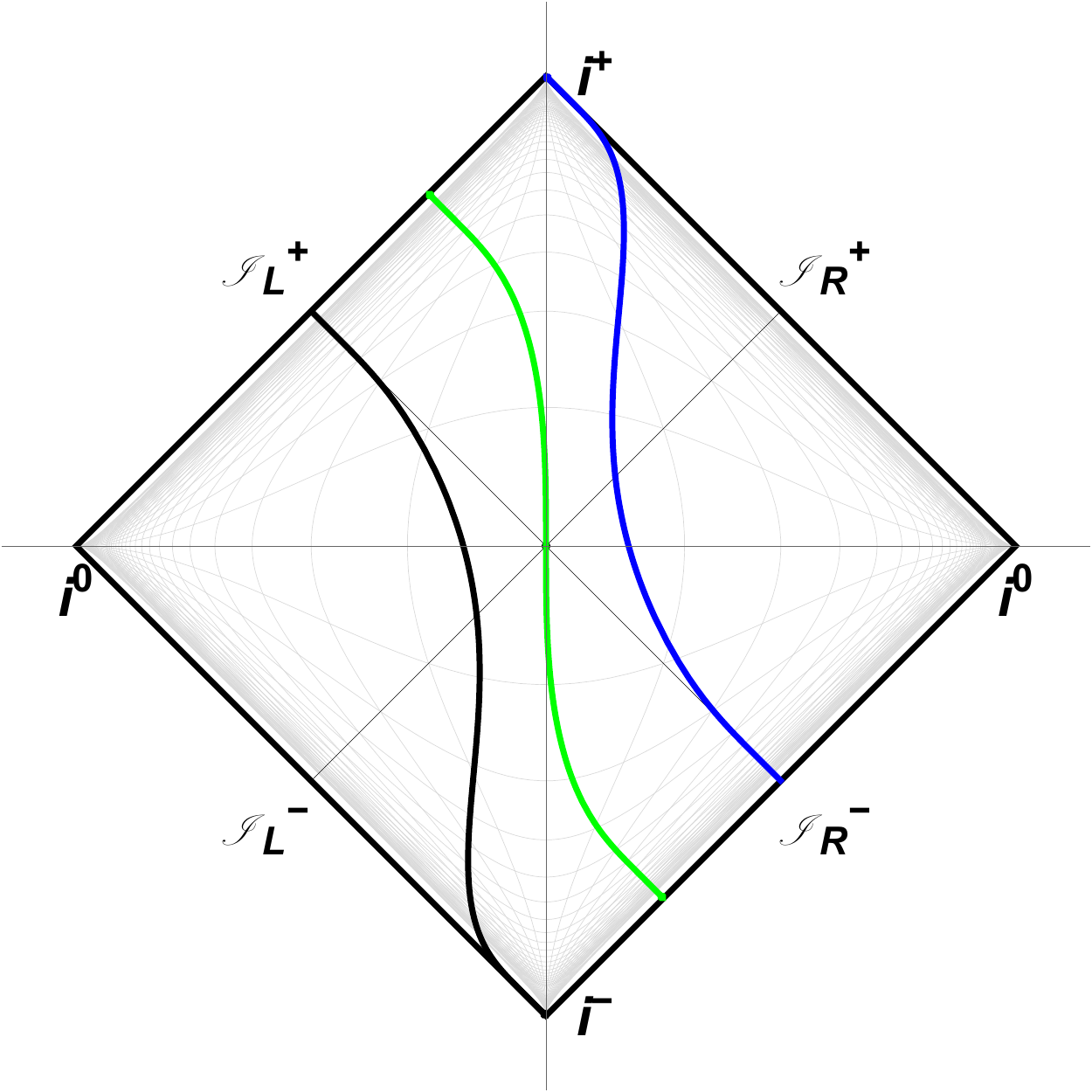} 
\caption{Penrose diagram of  constant energy flux mirrors with 
Planck distributed particles emission, showing Carlitz-Willey 
(black) Eq.~(\ref{eq:CW}), and a transformed case (blue), 
Eq.~(\ref{PT_CW}) with an equivalent spectrum. The de Sitter 
mirror (green), Eq.~(\ref{eq:ds}), is also related by M\"obius 
transform. Here $\kappa=2$ for illustration. }  
\label{fig1} 
\end{figure}

Figure~\ref{fig1} shows the trajectory as the solid black line, 
starting asymptotically inertial (no acceleration) albeit at  
light speed, and evolving dynamically toward asymptotic infinite 
acceleration with horizon at $v_H = 0$ in advanced time $v$. 
Since $p(u)$ is the function label for $v$, we see from 
Eq.~\eqref{eq:CW} that the mirror is limited to the bottom and 
left quadrants of the Penrose diamond.

Now let us apply the M\"obius transform, specifically the 
negative inversion of Eq.~\eqref{eq:pc} with $c=-b=1$, $a=d=0$. 
This gives 
\be 
P(u) = \frac{1}{\kp}\,e^{\kappa u}\,,  \label{PT_CW} 
\ee 
and is equivalent to the transform $x\to -x$, $t\to -t$, i.e.\ 
a $\mathcal{PT}$ symmetry. 
The trajectory is limited to the right and top quadrants of the Penrose 
diagram, plotted in Figure~\ref{fig1} as the solid blue line. 
The early time behavior illustrates the mirror climbing out of a horizon and approaching an asymptotic inertial end state at 
time-like future infinity. 
 
Explicit calculation indeed shows that the energy flux for both 
trajectories, as observed by a witness at future-null infinity, 
are identical, here the thermal constant, 
\be 
\mathcal{F} = \frac{\kp^2}{48\pi}\,. 
\ee 
One also finds that for the transformed case, like the original 
Carlitz-Willey mirror, the particles radiated are in a Planck 
distributed spectrum, 
\be 
N_{\omega\omega'} = \frac{1}{2\pi \kappa  \omega'}\,\frac{1}{e^{2\pi \omega/\kp}-1}\,, \label{orthodoxthermal}
\ee 
with temperature $T = \kp/2\pi$.
 
This result is not void of physical implications.  An observer who 
discovers constant energy flux emission and hot particles 
(with a temperature) will not be able to distinguish between the 
accelerated boundary conditions, nor determine the origin of such 
radiation.  Moreover, whether the horizon happened in the past, or 
in the future, cannot be determined.

To examine the particle creation in detail, we investigate the 
Bogolyubov beta function under the M\"obius transform. For 
this eternal thermal mirror, Eq.~(\ref{eq:CW}), the transform of 
$p(u)=-(1/\kp)e^{-\kp u}$ in the $c=0$ case is given by Eq.~\eqref{eq:pc0}; 
evaluating Eq.~\eqref{eq:bww} yields  
\be 
\beta_{\omega\omega'}\sim (-i\omega' a^2)^{-i\omega/\kp}\,\Gamma(i\omega/\kp)\,. \label{eq:betaetdil} 
\ee 
We now see how the extra constant $a$ from the (dilatation) 
transformation does not propagate to $\bsq$: 
\be 
(ix)^{iy}=x^{iy}\,e^{-(\pi/2)y\,sgn(x)}\, \label{eq:imagpow} 
\ee 
and so when multiplying it by its complex conjugate we simply 
get $e^{\pi y}=e^{-\pi\omega/\kp}$, independent of $x=-\omega' a^2$. Alternately, 
one can view $x^{iy}=e^{iy\ln x}$ and so $a^2$ enters as a pure 
phase. 

Next consider the eternal thermal $c\ne0$ case. For the 
$\mathcal{PT}$ transformed mirror, i.e.\ Eq.~\eqref{PT_CW}, 
the integral is very similar, giving 
\be 
\bww\sim (i\omega'/\kp)^{1-i\omega/\kp}\, \Gamma(1-i\omega/\kp)\,, 
\ee 
and this provides an identical $\bsq$ to that from 
Eq.~\eqref{eq:betaetdil} and hence the original Carlitz-Willey 
case: a thermal, Planckian particle spectrum. 

The more general case when $d\ne0$, i.e.\ Eq.~\eqref{eq:pc} with 
$s\ne0$, is considerably less tractable. The Bogolyubov beta function 
involves a Whittaker function, 
\be 
\bww\sim \Gamma\left(\frac{-i\omega}{\kp}\right)\,W_{i\omega/\kp,-1/2}(i\omega'/s)\,, \label{eq:gammaw} 
\ee 
but the spectrum per mode squared, $N_{\omega\omega'}$
\be \bsq = \frac{1}{4\pi\kp \omega' \sinh(\pi  \omega/\kp )} \left|U\left(\frac{-i \omega}{\kp} ,0,\frac{ i\omega '}{s}\right)\right|^2.\ee
is not the orthodox thermal spectrum, Eq.~(\ref{orthodoxthermal}), and thus $\bsq$ is not invariant for a general trajectory transformation. Here $U$ is the confluent hypergeometric function.  

When the argument of $W$ 
is large, e.g.\ the high frequency limit, then 
$W_{a,b}(z)\to e^{-z/2} z^a$. Since $z=i\omega'/s$ is imaginary, 
the first factor is a pure phase and cancels in the modulus, 
while $z^a$ gives a factor $e^{-\pi\omega/(2\kp)}$ using Eq.~\eqref{eq:imagpow}. 
Thus the $\bsq$ is the same as the original, Carlitz-Willey mirror, 
\be 
\bsq\sim \left|\Gamma\left(\frac{-i\omega}{\kp}\right)\right|^2\,e^{-\pi\omega/\kp}\sim \frac{1}{e^{2\pi\omega/\kp}-1}\,, 
\ee 
a Planckian particle distribution. Outside the high frequency regime the invariance of $\bsq$ is broken. 
However, at the level of the actual (observable) particle spectrum, 
$N(\omega)=\int  d\omega'\,\bsq$, 
the thermal spectrum is preserved. See Section~\ref{sec:nw} for 
the proof. 

This indicates that in a M\"obius transformation with $cd\ne0$, 
the particle production from the new mirror is not identical for 
the (unobservable) particle per mode per mode distribution. 
The low-frequency situation is reminiscent of the `soft' (zero energy) particle 
per mode per mode distribution of certain mirrors, such as in the 
uniformly accelerated case; we treat that in more detail in 
Section~\ref{sec:uniform}. 
However, the particles reaching a distant observer would be blueshifted, and therefore in the high frequency limit where the per mode per mode distribution is invariant. And as mentioned in the previous 
paragraph, the observable particle spectrum $N(\omega)$ is thermal and invariant.

\section{de Sitter} \label{sec:dS} 

A de Sitter space has a horizon and eternally thermal emission; the 
accelerated boundary correspondence for it is a mirror with 
trajectory  \cite{Good:2020byh}
\be 
p(u)=\frac{2}{\kp}\,\tanh\frac{\kp u}{2}\,. \qquad({\rm de\ Sitter}) \label{eq:ds} 
\ee 
Using the M\"obius transformations we can convert this into 
the Carlitz-Willey (CW) eternal thermal mirror of Eq.~\eqref{eq:CW}, 
or the reverse: 
\bea 
P_{\rm dS}&=&\frac{-1}{c^2p_{\rm CW}+s}\qquad {\rm with}\ c^2=\frac{\kp}{4}=-s,\\ 
P_{\rm CW}&=&\frac{-1}{c^2p_{\rm dS}+s}\qquad {\rm with}\ c^2=\frac{\kp}{4},\ s=\frac{1}{2}\,.  
\eea  
As before we have dropped constant contributions to $P$ as they only give 
a constant phase (though they do affect the $v$ location 
of the horizon). See Figure~\ref{fig1} for a plot of the de Sitter mirror (green), Eq.~(\ref{eq:ds}) with $\kappa = 2$.  

We can use the same transformations for anti-de Sitter space, 
where $p=(2/\kp)\tan(\kp u/2)$, 
as long as we take $\kp_{\rm AdS}=\pm i\kp_{\rm dS}$, since AdS 
has negative eternal thermal 
flux $F=-\kp^2_{\rm dS}/(48\pi)$. 

As for the eternal thermal mirror with $d\ne0$ (recall $s=cd$), 
the Bogolyubov beta coefficient resolves to Eq.~\eqref{eq:gammaw}, which by converting to the confluent hypergeometric function $M$ is equivalent to Eq.~(16) of \cite{Good:2020byh}, 
and has the same `soft' particles in $\bsq$, while keeping an invariant thermal particle spectrum $N(\omega)$.

\section{Schwarzschild and Kerr} \label{sec:schw} 

The Schwarzschild ABC has mirror trajectory \cite{Good:2016oey}
\be 
p(u) = v_H - \frac{1}{\kappa}\, W\left(e^{-\kappa(u-v_H)}\right)\,, \label{Schwarschildmirror}
\ee 
where $W$ is the Lambert $W$ function, also known as a product log. 
Calculating the particle spectrum of this under M\"obius transforms 
is not analytically tractable, so we make some general observations. 

\begin{figure}[h]
\centering 
\includegraphics[width=3.0in]{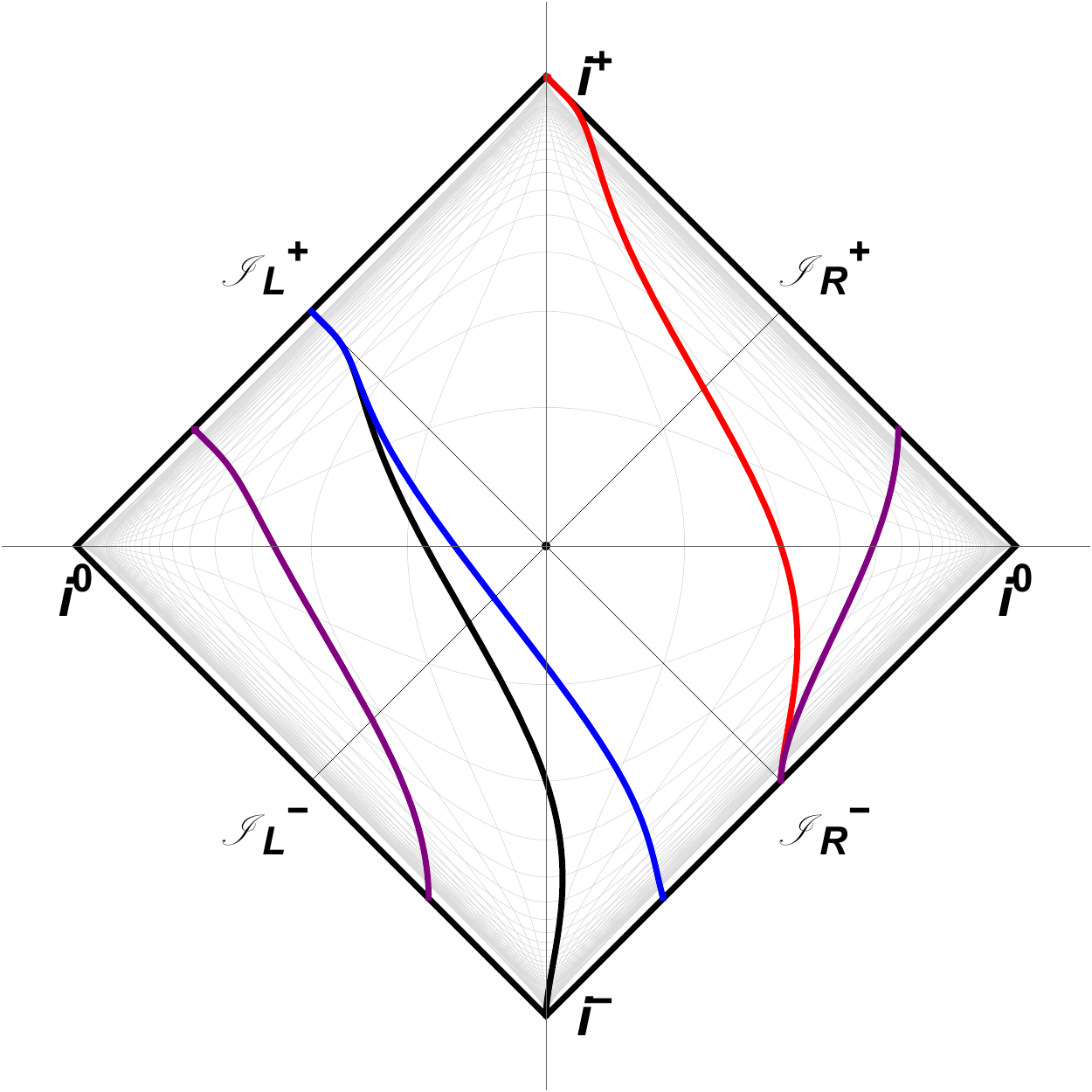} 
\caption{Penrose diagram of some identical flux mirrors.  The original Schwarzschild mirror, Eq.~(\ref{Schwarschildmirror}), is black. The colors, (red, blue, purple), correspond to $P(u)$ with schematic form: $1/W$,  $-W/(W+1)$, $1/(W-1)$, respectively. The purple line (both branches are shown) is a M\"obius transform of the Schwarzschild mirror with $u$-horizon at $u=-1$, where the mirror, Eq.~(\ref{midhorizon}), runs into the observer at right null-infinity.  The energy flux detects nothing unusual. 
} 
\label{fig3} 
\end{figure}

Figure~\ref{fig3} plots the Schwarzschild mirror trajectory as the 
black curve. The other curves in the figure show the application of 
M\"obius transforms, and one can verify that they have identical 
energy flux to the original Schwarzschild mirror. (In fact, one can 
even create spacelike trajectories with the same energy flux.)

Note that one of the M\"obius transforms, with $a=0, d=c=1, b=-1$, 
of the Schwarzschild mirror (with $\kappa = 1$, $v_H=0$ for simplicity) 
gives the purple curve, 
\be 
P(u) = \frac{1}{W\left(e^{-u}\right)-1}   \label{midhorizon} 
\ee 
with identical energy flux and a horizon in $u$, but where the mirror 
smashes into our intrepid observer at null-infinity. Despite this, 
the witness will register nothing unusual happening in the energy flux! 
Interestingly, one can imagine wrapping the 
space so the points $(u,v)=(-1,+\infty)$ and 
$(-1,-\infty)$ join, making the purple mirror trajectory 
continuous (speculatively, a wormhole?). We consider an implication of this sort of 
idea in Sec.~\ref{sec:uniwrap}. 

In the hopes of dealing with simpler expressions than product logs, 
and exploring the Kerr metric for which $p(u)$ is unknown, one might use the label function $f(v)$ describing the null 
coordinate $u$. For Kerr (and Schwarzschild as a limiting case), 
\be 
f(v)=v-\frac{1}{2g}\frac{1+\delta}{\delta}\,\ln|gv|+\frac{1}{2g}\frac{1-\delta}{\delta}\,\ln|gv-\delta|\,, \label{eq:fkerr} 
\ee 
where $g=1/(4M)$ and $\delta=\sqrt{1-a^2/M^2}$, with $M$ and $a$ 
the mass and spin parameter of the black hole. 
For the Schwarzschild case $a=0$, $\delta=1$, $g=\kappa$, and the 
last term in $f(v)$ vanishes. 
However, the M\"obius transformation only applies 
to $p(u)$. 
See Appendix~\ref{sec:fv} for discussion 
of transformations applied to $f(v)$.

\section{Constant Acceleration Mirrors} \label{sec:uniform} 

Mirrors with constant acceleration have zero energy flux. However, 
the particle production is a more subtle matter.  Let us explore 
what M\"obius transformations of uniform acceleration look like. 
(Of course they will preserve the zero energy flux.) 
The classic uniform acceleration mirror trajectory is given by $p(u)=-1/(\kp^2 u)$, 
with acceleration $\alpha(u)=p''(u)/[2(p')^{3/2}]=-\kp$.

\subsection{Branches Matter} 

\begin{figure}[h]
\centering 
\includegraphics[width=3.0in]{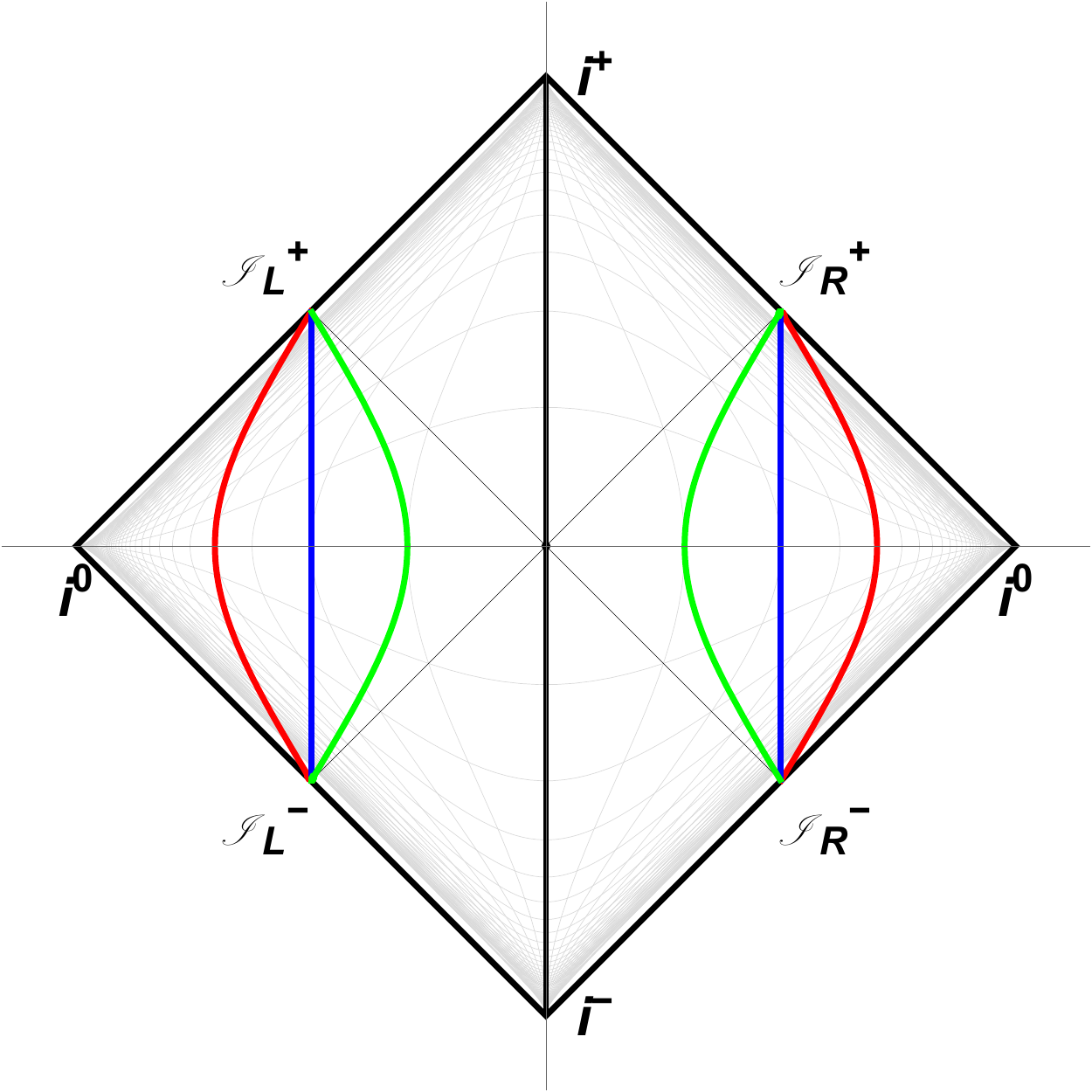} 
\caption{Penrose diagram of the uniformly accelerated mirror, with $\kappa = 1/2, 1, 2$, for red, blue, green, respectively.  Note that $\kappa =1$ (blue) is a straight line on the conformal diagram but is not stationary in space. Horizons exist at $u=0$ and at $v=0$.  
The particle production and energy emission is zero. }  
\label{fig2} 
\end{figure}

If we M\"obius transform with $c=0$, then $P(u)=a^2p=-a^2/(\kp^2 u)$. 
This dilatation is merely a redefinition of $\kappa$, i.e.\ a 
different constant acceleration. As mentioned, the energy flux 
remains zero, but the particle production involves a subtlety: 
there are two branches, $u<0$ and $u>0$ on either side of the 
horizon at $u=0$. These can be viewed as two mirrors. 
A Penrose diagram of the dual-mirror system situation is given in 
Figure~\ref{fig2} depicting the two mirrors (each shown 
for various $\kappa$ parameters) that are the two roots of the 
single spacetime coordinate trajectory function. In spacetime 
coordinates, $x$ and $t$, hyperbolic motion is 
\be 
x^2 - t^2 = \alpha^{-2}\,, 
\ee
so that there are actually two mirrors: 
\be 
x(t) = \pm \sqrt{\alpha^{-2} + t^2}\,, 
\ee 
where $\alpha = -\kappa$ is the constant proper acceleration for each. 
The mirror in the right quadrant of the diagram traverses $(u,v)=(-\infty,0)$ to $(0,\infty)$ while the one in the left 
quadrant traverses $(0,-\infty)$ to $(\infty,0)$.

\subsection{Zero Particle Creation with Two Branches} \label{sec:uniwrap} 

For a $c\ne0$ M\"obius transform, 
\be 
P(u)=\frac{-1}{-c^2/(\kp^2 u)+s}=-\frac{1}{s}-\frac{c^2}{s(s\kp^2u-c^2)}\,, 
\ee 
and we can ignore the first term as contributing a constant phase. 
Switching to $U=s\kp^2u-c^2$, we have $P(U)=-c^2/(sU)$. 
The integration in $\bww$, split into $[-\infty,0]$ and $[0,\infty]$, 
{\it but counting both branches\/} 
gives 
\be 
\bww\sim \sqrt{\frac{4q}{\omega}}\,e^{i\pi/2}\, 
\left[K_{-1}\left(-2\sqrt{q\omega}\right)+K_{-1}\left(2\sqrt{q\omega}\right)\right]\,, 
\ee 
where $q=\omega'/\kp^2$ for $p(u)$ and $q=(c^2/s^2)\omega'/\kp^2$ 
for the transformed $P(u)$. Since 
for the modified Bessel function of the second kind $K_{-1}(-z)=-K_{-1}(z)$, we 
find $\bww=0$ regardless of $q$ and hence $c$, $s$. 

Thus, when taking into account both branches, the transformation leads to both invariant (zero) flux and (zero) particle 
production for the constant acceleration case.

\subsection{Non-zero Particle Creation with One Branch}

In the previous subsection we worked with a dual mirror system (two roots) (see also \cite{Fabbri}).  Now we will look at just the single mirror system (one root) with uniform acceleration, which has long been known to have the surprising result of non-zero particle production with zero energy flux \cite{Birrell:1982ix}. To compute the soft particle production from a single uniformly accelerating mirror, we can shift it over to the origin to avoid confusion with the previous system, expressing the single uniformly accelerated mirror, again with $\alpha = -\kappa$, as
\be 
p(u) = \frac{u}{1+\kappa u}\,, 
\ee 
with early-time horizon positioned at $u_H = -\kappa^{-1}$.  It has particle creation that is solved by integrating from the early time horizon onward (this branch only includes one root, and therefore only one mirror) via 
\be 
\beta_{\omega\omega'} = \frac{-1}{2\pi} \sqrt{\frac{\omega}{\omega'}}\int_{-\kappa^{-1}}^\infty du\, e^{-i \omega u -i \omega' u/(1+\kp u)}\,.  
\ee 
Using a simple substitution $X=u+1/\kappa$ gives a range of integration from $[0,\infty]$ and  
\be 
\beta_{\omega\omega'} = \frac{-i}{\pi\kappa}\, e^{i (\omega-\omega')/\kappa}\, K_1\left(\frac{2}{\kappa} \sqrt{\omega \omega'}\right)\,. 
\ee
The phase pre-factor is unimportant (as well as the sign of the beta) upon complex conjugation, so 
\be 
|\beta_{\omega\omega'}|^2 = \frac{1}{\pi^2\kappa^2} \left| K_1\left(\frac{2}{\kappa} \sqrt{\omega  \omega '}\right)\right|^2\,.\label{UAbeta2} 
\ee
Eq.~(\ref{UAbeta2}) is the `soft'-spectrum per mode per mode of a single uniformly accelerated mirror, distinctly non-thermal (non-Planckian) \cite{Davies:1976hi, Davies:1977yv, Birrell:1982ix}; in the high frequency limit $K_1\to0$ (exponentially), however, 
and as before the distribution is independent 
of transformation.  
Thus, while the union of the two branches preserves the particle flux per mode per mode (zero, for a uniform acceleration) under a M\"obius transform, a single branch does not, except in the high frequency limit. However, as we shall see in Section \ref{sec:nw}, the particle spectrum  $N(\omega)$ is preserved.

\subsection{Thermal Uniform Acceleration} \label{HotUniform}

While a single uniformly accelerated mirror produces soft particles, despite zero energy flux, and we have seen that the spectrum is decidedly non-Planckian \cite{Birrell:1982ix}, the story of the radiation is not told by just the acceleration. Is there a context where a uniformly accelerated moving mirror can have a temperature?  Here we find that yes, it almost certainly can, under very specific circumstances.  This is similar to the Unruh effect \cite{unruh1976} where an eternally uniformly accelerated observer (not mirror) sees a Planck distributed particle radiation with temperature proportional to the acceleration.  The situation we will investigate will be in stark contrast to the original situation of the Davies-Fulling effect for a single moving mirror that eternally uniformly accelerates creating a Bessel distributed particle radiation distribution per mode per mode, Eq.~(\ref{UAbeta2}), with undefined temperature and zero energy flux (soft particles).  If we relax the eternal uniformity, and look at what happens as a mirror {\it asymptotically\/} approaches uniform acceleration, we find a particular mirror that can emit constant energy flux indicative of thermal emission.

\begin{figure}[h]
\centering 
\includegraphics[width=3.0in]{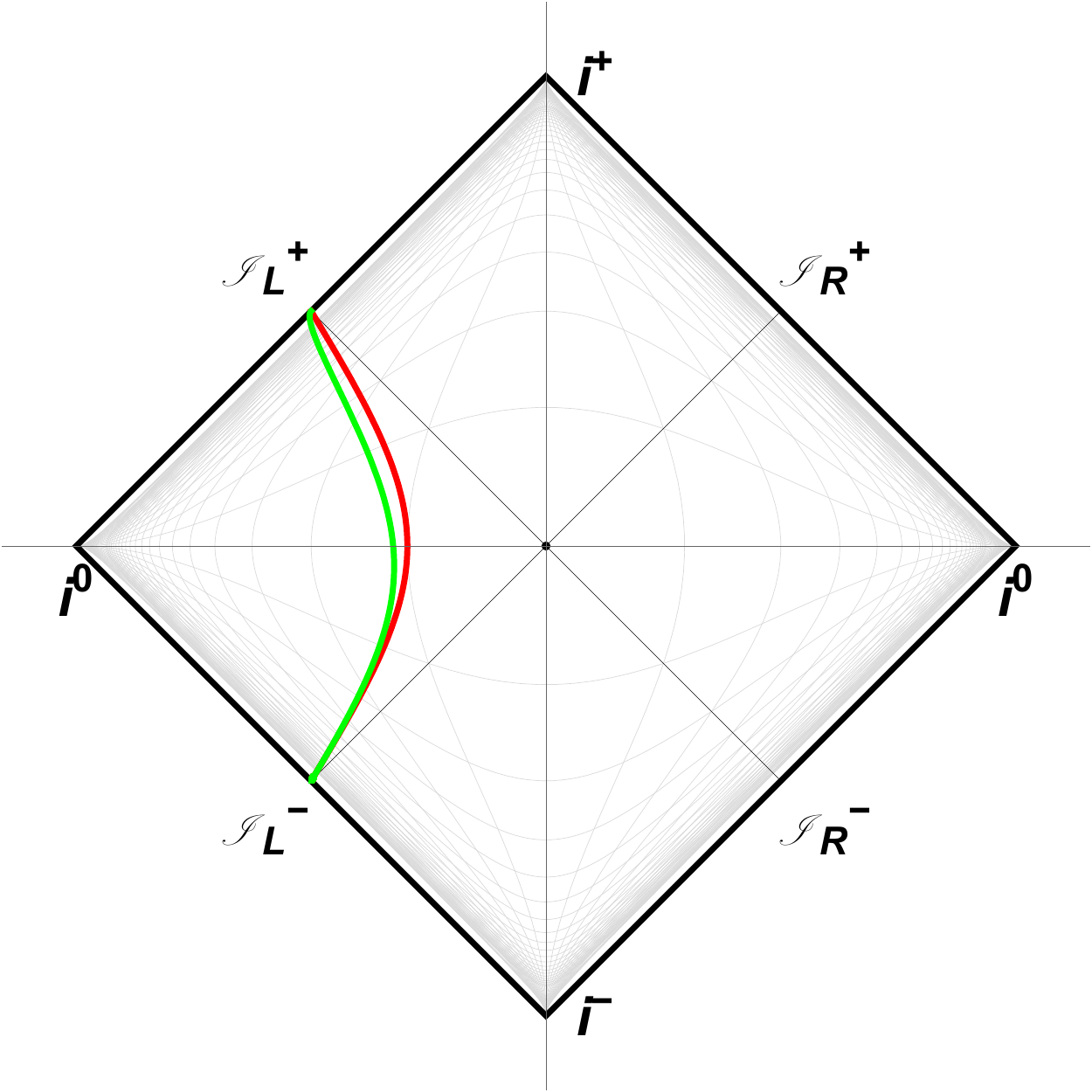} 
\caption{Penrose diagram of a uniformly accelerated mirror (red), $f(v) = -1/(\kappa^2v$), (branch $v<0$ only), and an early-time asymptotically uniformly accelerated mirror (green), Eq.~(\ref{earlyUA}).  Here $\kappa = 2$ for illustration.}  
\label{fig2a} 
\end{figure}

From the start of this section, recall that $p(u)=-1/(\kp^2u)$ 
has eternal uniform acceleration. This can be written in terms 
of $f(v)=-1/(\kp^2 v)$. Now consider the trajectory 
\be 
  f(v)=-\frac{1}{\kp^2 v}-\frac{1}{12\kp^4 v^3}\,. \label{earlyUA} 
\ee 
(One could obtain $p(u)$ by solving the cubic 
equation, but dealing with $f(v)$ is simpler.) 
This asymptotically approaches the uniform acceleration 
trajectory as $v\to-\infty$. The proper acceleration 
$\alpha(v) = -f''(v)/[2 f'(v)^{3/2}]$ has an early time limit 
\be 
\lim_{v\to -\infty} \alpha(v) = 
 -\kappa\,\left[1+\mathcal{O}\left((\kp v)^{-2}\right)\right]\,, 
\ee 
explicitly approaching uniform acceleration (accelerating leftward using the Davies-Fulling convention \cite{Davies:1976hi, Davies:1977yv}). 
The trajectory (green) is shown in Figure~\ref{fig2a}, along with the 
eternally uniform acceleration case (red).

The energy flux as derived by the quantum stress tensor for the moving mirror model is also easily found in advanced time using the Schwarzian, $24\pi \mathcal{F}(v) = \{f(v),v\}f'(v)^{-2}$, which has an early time limit, 
\be 
\lim_{v\rightarrow -\infty} \mathcal{F}(v) = \frac{\kappa^2}{48\pi}\,. \label{PEFconstant} 
\ee 
This demonstrates conclusively that despite (asymptotically) uniformly accelerating, the energy flux emission is a non-zero constant, unlike the eternally uniformly accelerating mirror \cite{Davies:1976hi}. 
The key here is that $\alpha'(v)$ is not exactly zero, although 
it is asymptotically zero. 
The constant energy flux of Eq.~(\ref{PEFconstant}) is identical to that  of the de Sitter mirror \cite{Good:2020byh} or the eternal thermal mirror of Carlitz-Willey \cite{carlitz1987reflections}, which have  explicitly been shown to emit particles in a Planck distribution with temperature 
\be 
T=\frac{\kappa}{2\pi}\,. 
\ee 

We study approaches to asymptotes further in Appendix~\ref{sec:fv}.

\section{Particle Spectrum Invariance} \label{sec:nw} 

In the previous sections we have concentrated on the 
energy flux and the particle spectrum per mode per mode, 
i.e.\ $N_{\omega\omega'}=\bsq$. Concentrating on the observable particle 
spectrum itself, 
\be 
N(\omega) = \int_0^\infty d\omega'\, |\beta_{\omega\omega'}|^2\,, 
\ee 
its invariance under M\"obius transformations 
can be shown quite 
generally\footnote{In a late stage of 
writing we found an argument by 
\cite{Reuter:1988nt}, of which ours is 
an inverse; we are indebted to it for  
improving the derivation to that 
given here.}. 

To do so, it is most convenient to work with $f(v)$, i.e.\ 
the function label for the retarded time $u$. Here the 
Bogolyubov beta coefficient takes the form 
\be 
\beta_{\omega\omega'} = \frac{1}{2\pi}\sqrt{\frac{\omega'}{\omega}}\int_{v_{\textrm{min}}}^{v_{\textrm{max}}} dv\, e^{-i \omega' v - i\omega f(v)}\,. 
\ee 
In computing the particle spectrum, the proof proceeds by 
carrying out the 
integration over $\omega'$ first, so that 
\bea 
N(\omega)&=&\frac{1}{4\pi^2\omega}\int dv_1\int dv_2\,e^{-i\omega [f(v_1)-f(v_2)]}\notag\\ 
&\qquad&\times \int d\omega'\,\omega'\,e^{-i\omega'(v_1-v_2)}\notag\\ 
&=&\frac{-1}{4\pi^2\omega}\int\int 
\frac{dv_1 dv_2}{(v_1-v_2-i\eps)^2}\,e^{-i\omega [f(v_1)-f(v_2)]}\,, \label{eq:Nw1}
\eea 
where we used a real regulator $\eps>0$ that we then take the 
limit $\eps\to0$. 

Now consider another mirror trajectory that incorporates the 
M\"obius transform of $p(u)$, i.e.\ the function label for $v$. 
We call $V_i=Mv_i$. Its particle spectrum will simply be 
\be 
\tilde N(\omega)=\frac{-1}{4\pi^2\omega}\int\int 
\frac{dV_1 dV_2}{(V_1-V_2-i\eps)^2}\,e^{-i\omega [\tilde f(V_1)-\tilde f(V_2)]}\,, \label{eq:Nw2} 
\ee 
where we indicate this mirror's quantities with tildes. 

However, a property of the M\"obius transform is that the quantity 
\be 
\frac{P'(u_1)P'(u_2)}{[P(u_1)-P(u_2)]^2}= 
\frac{p'(u_1)p'(u_2)}{[p(u_1)-p(u_2)]^2}\, \label{eq:jacob} 
\ee 
is invariant. Since $p(u)$ is the function label for $v$ 
(and the same relation of $P(u)$ with $V$), this is 
precisely the integration ``Jacobian'' 
$dV_1 dV_2/(V_1-V_2)^2$. Furthermore, since $f(v)$ is the label 
function for $u$, and $u$ is kept fixed during the transform 
$p(u)\to P(u)$, then $\tilde f(V=Mv)=f(v)$. This shows that 
Eq.~\eqref{eq:Nw1} is identical to Eq.~\eqref{eq:Nw2}, and 
hence that the particle spectrum 
\be 
\tilde N(\omega)=N(\omega)\,,  
\ee 
is invariant between M\"obius transformed mirrors. 

In summary, the M\"obius transformation preserves the 
energy flux and the particle spectrum, though the Bogolyubov 
beta coefficients can differ.  The particle distribution 
per mode per mode can also deviate due to soft particles 
when $cd\ne0$, but is preserved in the observable 
(high frequency) limit.

\section{Conclusions} \label{sec:concl} 

The anomalous quantum effect that breaks conformal symmetry of classical relativity is particularly explicit and manifest in the symmetries of the moving mirror model and its resulting radiation from vacuum.  Although gravitation does not extend to the (1+1)-dimensional spacetime of the model (where $G_{\mu\nu}$ is identically zero), and the full physical significance of the gravitational analog in the ABC is limited to the strength of the analogy between (1+1) and (3+1)-dimensions, the tractability of fully analytical results due to the additional simplicity gained by avoiding spacetime curvature is an asset. As far as quantum fields and relativistic thermodynamics are concerned this approach surrenders some valuable albeit  phenomenological insights. 

Our work underscores the symmetry governing the conformal anomaly of the moving mirror model.  The most fundamental observables of the model are invariant under M\"obius transformations.  This includes the energy flux $\mathcal{F}$, the particle spectrum $N(\omega)$, and the von Neumann entanglement (geometric) entropy $S_\textrm{ren}$. In select important cases we have found that this invariance extends to the spectral measure $N_{\omega\omega'} =\bsq$.
We summarize our results: 

\begin{itemize}
    \item Explicit demonstration of the dramatic dynamical change that a M\"obius transform inflicts on the mirror trajectory, while not impacting what the observer sees. Humorously phrased, this gives a `Mirror of Dorian Gray', or 
as seen in Figure~\ref{fig1}, `objects in moving mirror are closer (or further) than they appear'; see start of  Section \ref{sec:eternal}. 
    \item Full particle invariance $\bsq$ for the eternal thermal mirror under dilatations or inversions ($cd=0$ case). However, $\bsq$ can differ for transforms where $cd\ne0$, 
though invariance is restored in the high frequency regime, and 
upon integration over the ingoing modes to form the observable particle spectrum $N(\omega)$; see end of Section~\ref{sec:eternal} 
and Section~\ref{sec:nw}. 
    \item Explicit M\"obius transforms between the 
Carlitz-Willey mirror and de Sitter mirror. Although both are 
eternally thermal, the Bogolyubov beta coefficients and $\bsq$ 
are not invariant under the transforms, due to changing horizon structure; see Section~\ref{sec:dS}. 
    \item A M\"obius transform on the 
Schwarzschild mirror can give a horizon unseen by the stress tensor, Eq.~(\ref{midhorizon}) and Figure~\ref{fig3}, demonstrating that the energy flux may contain hidden horizons, or a wrapping of the spacetime itself; see Section~\ref{sec:schw}. 
    \item Consistency of zero energy and zero particles for dual uniform accelerating mirrors; 
see Section~\ref{sec:uniwrap}.
    \item Shown how asymptotic uniform acceleration can radiate thermal energy flux; see Section~\ref{HotUniform}. 
    \item Shown that zero total energy time-dependent accelerating mirrors are forbidden; see Appendix~\ref{sec:zeroen}. 
\item Elucidation of invariance from M\"obius transformations for entanglement entropy; see Appendix~\ref{sec:Entropy}. 
    \end{itemize} 

The moving mirror model has come a long way over the five decades since DeWitt \cite{DeWitt:1975ys}, Davies and Fulling \cite{Davies:1976hi,Davies:1977yv} demonstrated accelerated boundaries produced particles from the quantum vacuum. 
It has flourished, with e.g.\ Schwinger \cite{Schwinger4091}, Unruh \cite{Davies:1976ei}, Wilczek \cite{wilczek1993quantum} 
helping to establish theoretical and experimental \cite{Chen:2020sir,Chen:2015bcg} research directions into 
the nature of particle creation, spacetime, and information.  
In 2021 alone, the dynamical Casimir effect field has been 
enriched by diverse, novel ideas ranging over holography,  
qubits, relativity, and harvesting entropy  
\cite{Liu,Reyes,Akal:2020twv,Akal:2021foz,Lin:2021bpe,Kawabata:2021hac,Ageev:2021ipd,Agusti:2020zro,Alves:2020vgl,Sato:2021ftf}.  
The connection of the M\"obius transform to the group 
$SL(2,\mathbb{R})$, Schwarzian derivatives, and conformal 
field theory offers another avenue to a deeper understanding 
of the nature of information, horizons, and vacuum particle creation.

\acknowledgments 

MG acknowledges funding from state-targeted program ``Center of Excellence for Fundamental and Applied Physics" (BR05236454) by the Ministry of Education and Science of the Republic of Kazakhstan, and the FY2021-SGP-1-STMM Faculty Development Competitive Research Grant No.\ 021220FD3951 at Nazarbayev University.
This work is supported in part by the Energetic Cosmos Laboratory. EL is supported in part by the U.S.\ Department of Energy, Office of Science, Office of High Energy Physics, under contract no.\ DE-AC02-05CH11231.

\appendix

\section{Transforming $f(v)$} \label{sec:fv} 

As seen in the Schwarzschild case, the trajectory $p(u)$ 
in terms of the null coordinate $u$ is not always 
tractable to general manipulation. It is worthwhile 
exploring what the equivalent of the M\"obius transformation 
is for $f(v)$. That is, what mapping $f(v)\to \tilde{f}(V=Mv)\equiv g(v)$ keeps the 
energy flux invariant. Recall that 
\be 
24\pi \mathcal{F}(v)=\frac{1}{f'(v)^2}\left[\frac{f'''}{f'}-\frac{3}{2}\left(\frac{f''}{f'}\right)^2\right]\,. 
\ee 

We write $g'(v)=h(f)f'(v)$ to obtain a differential equation for 
$h(f)$. Denoting $\dot h=dh/df$, we have 
\be 
\ddot h-\frac{3}{2}\frac{\dot h^2}{h}+(h-h^3)\,24\pi \mathcal{F}(f)=0\,, \label{eq:ftrans} 
\ee 
where $\mathcal{F}(f)$ is the energy flux we seek to keep invariant. 

Consider the case of zero flux, i.e.\ the constant acceleration  
mirror of Sec.~\ref{sec:uniform}. Here the mapping resolves to 
\bea 
f(v)&\to& \tilde{f}(V=Mv)\equiv g(v)\notag\\ 
&\qquad&=k_1+k_2^{-2}f(v)\,\delta(k_3)-\frac{\cancel{\delta}(k_3)}{k_3^2f+k_2k_3}\,, \label{eq:gzero} 
\eea 
where $k_1,k_2,k_3$ are constants. The $\delta$ notation indicates that 
the second term only exists if $k_3=0$ (and $k_2\ne0$) and the third 
term only exists if $k_3\ne0$ (so one never has both terms at once). 
We indicate that when we apply to a M\"obius 
transform to $v$, written as $V=Mv$, then the 
function label for $u$  goes from $f(v)$ to $\tilde f(V)$. 
The trajectory $f(v)=-1/(\kp^2 v)$ gives 
the constant acceleration, zero energy flux case, so our results 
indicate that any mirror with trajectory 
\be 
g(v)=k_1-\frac{\delta(k_3)}{k_2^2\kp^2v}-\frac{\cancel{\delta}(k_3)}{k_2k_3-k_3^2/(\kp^2v)}\,, \label{eq:guni} 
\ee 
will also have zero energy flux. The case with $k_3=0$ is  
a dilatation, 
a redefinition of $\kp$. When $k_2=0$, $k_3\ne0$, then $g=\kp^2v/k_3$ and 
this is an inversion, to the trivial constant velocity, zero acceleration case. 
The case with $k_3\ne0$, $k_2\ne0$ has 
$g=-\kp^2v/(k_2k_3\kp^2v-k_3^2)$, and the mirror has $g(v=\pm\infty)=-1/(k_2k_3)\ne0$, 
$g(0^\pm)=0^\pm$, unlike $f(v=\pm\infty)=0^\mp$, $f(0^\pm)=\mp\infty$. 

When the flux is instead thermal, and hence constant, we 
can solve Eq.~\eqref{eq:ftrans} to find 
\be 
f(v)\to g(v)=k_1-\frac{1}{\kp}\,\ln\left(\frac{e^{\kp f}-1}{e^{\kp f}+1}\right)\,. \label{eq:gtherm} 
\ee 
One can verify that Carlitz-Willey $f(v)$ can be transformed 
into de Sitter, and vice versa, with this formula. 

It is also interesting to consider when two ABC have only 
asymptotically the same energy flux, whether thermal or 
zero. 
The transformation teaches us that the approach to 
thermality is nearly identical for the transformed trajectory 
relative to the original thermal mirror, even if it is only 
thermal asymptotically. 
Consider $g'(v)=[1+\eps(v)]\,f'(v)$. Then Eq.~\eqref{eq:ftrans} 
becomes $\ddot\eps=48\pi \mathcal{F}\eps$ to first order, and 
\be 
\eps(f)\sim e^{-\int^f_{f_0} df_\star\,\sqrt{48\pi \mathcal{F}(f_\star)}}\,, 
\ee 
as long as the $\dot h^2/h\sim \dot\eps^2$ term is small compared 
to the leading order. 
Approaching thermality, $\mathcal{F}\to \mathcal{F}_{\rm th}=\kp^2/(48\pi)$, and 
the distance moved $\Delta f\to\infty$ as seen from Eq.~\eqref{eq:fkerr}, 
or equally the eternal thermal or de Sitter cases, 
so indeed 
$g'(v)\to f'(v)$ exponentially. Since the energy flux depends only 
on $f'$ (and its derivatives), then to leading order the transformed and original mirror 
trajectories approach thermality in the same way. 

Similarly, when the flux approaches zero, then on that asymptote 
we can map the 
uniform acceleration mirror to, for example, Schwarzschild, which 
has $\mathcal{F}(v\to-\infty)\sim v^{-3}\to0$. This can be done through 
the transformation of Eq.~\eqref{eq:guni} with $k_2=0$, $k_3=\kp$, 
giving the leading order in 
$f_{\rm Schw}(v)=v-\kp^{-1}\ln(-\kp v)\approx v$ as $v\to -\infty$. 
Unfortunately, using the full Schwarzschild flux 
\be 
\mathcal{F}_{\rm Schw}(v)=\frac{\kp^2}{48\pi}\frac{1-\kp v}{(1-\kp v)^4}\,, 
\ee 
in Eq.~\eqref{eq:ftrans} is not tractable due to $\mathcal{F}(f)$ 
leading us back to product logs.

\section{No Go Zero Energy} \label{sec:zeroen} 

We have explored transformations that leave invariant the energy flux. 
If the energy flux is unchanged then of course the total energy 
emitted is as well. Suppose we now consider mirrors with the same 
total energy, without requiring the energy flux $\mathcal{F}(u)$ be 
the same. For mirrors where the total energy is infinite (e.g.\ 
the eternal thermal mirrors), this question is not so interesting: 
there are many ways to add up to infinity. The other total energy 
of particular interest is zero net energy, such as seen in the 
constant acceleration case. There, the zero energy arises from 
uniform zero energy flux.  However, we know that 
negative energy flux can exist as well as positive energy flux, and 
indeed is required under certain widespread circumstances, such as 
unitarity \cite{Good:2019tnf,Bianchi:2014vea,Bianchi:2014qua}.  

Can we arrange mirror motion such that negative energy flux at some time exactly 
cancels positive energy flux at some other time, leaving zero total energy? We present  
here a no-go conjecture against this possibility. What we find is 
that such a trajectory balancing the negative and positive energy 
fluxes is always interrupted by a horizon, leaving a net positive 
or negative total energy. 

It is useful to write the energy flux in terms of the rapidity  
$\eta(u)=(1/2)\ln p'(u)$. Then the total energy is 
\be 
E=\int_{u_-}^{u_+} du\,\left[\eta''-\left(\eta'\right)^2\right]\,,  
\ee 
where $u_-$, $u_+$ are the limits of mirror motion, either 
finite or infinite horizons. 

The first term is a total derivative, so 
\be 
E=\eta'(u_+)-\eta'(u_-) - \int du\,\left(\eta'\right)^2\,. 
\ee 
To achieve zero energy, since the remaining integrand is 
positive we require $\eta'_+>\eta'_-$. 

Now if $\eta'(u_+)\to\pm\infty$ or $\eta'(u_-)\to\pm\infty$ then 
the second term, involving $\eta'$, must win, giving negative total 
energy. If $\eta'\to$ const ($\ne0$), then if $u$ extends to 
$\infty$ again the integral will dominate the boundary term, 
since the sum of a near constant over an infinite interval 
is infinite. Thus we need either $\eta'\to 0$ (asymptotic inertia) 
or to cut off $u$ by a horizon at finite coordinate. 

A horizon has $\eta=(1/2)\ln p'\to\infty$, so there is a pole in 
$p'(u)$. Noting that $\eta'=p''/(2p')$, we would normally expect 
$\eta'\to\infty$, i.e.\ if the pole is of order $m$ so $p'\sim 1/(u_+-u)^m$ 
then $\eta'=m/[2(u_+-u)]\to\infty$ as $u\to u_+$ (the same holds for obtaining 
the horizon with $\eta\to\infty$ by $p'\sim (u-u_\star)^m\to0$, 
i.e.\ the previous $m$ is negative). Hence this reduces to the 
previously considered case, which did not give zero flux. 
However, there are special cases where $p''\sim p'$ 
($\eta'\to{\rm const}$) and $p''/p'\to0$ ($\eta'\to0$). The first 
of these has $p'\sim e^{bu}$, which blows up at $u_+=\infty$; this 
is not at finite $u$ and this is the previously considered, failed 
case  of $\eta'\to{\rm const}$. The second of these takes 
$\eta'\to0$ and is our remaining case to assess. 

For this last case, where $\eta'\to0$, then unless $\eta'=0$ for 
all $u$, the range where it is nonzero will contribute to the 
integral and again unbalance the terms, preventing zero energy. 
The only zero total energy case is the zero energy flux case, 
when $\eta''=(\eta')^2$, and hence $\eta'=-1/u$. This is 
precisely the constant acceleration case with $p(u)=-1/(\kp^2u)$. 
This case does have a pole, at $u=0$, and $\eta'\to\infty$ there,  
but the terms balance in this one case. For poles of order $m$, 
the integral gives $\int du\,u^{-2m}=(1-2m)^{-1} u^{-2m+1}$ while the boundary 
term gives $u^{-m}$. Only when $-2m+1=-m$, i.e.\ the $m=1$ case, 
do the terms cancel each other.

\section{Entanglement Invariance}\label{sec:Entropy}

Here we derive the (1+1)-dimensional entanglement entropy or `geometric entropy' in conformal field theory (CFT) and show its fundamental invariance under M\"obius transformations of the trajectory. We then break the invariance by referencing a static mirror on one side of the system, which amounts to a choice of a reference frame for the observer, deriving the relationship to the rapidity of the mirror (see e.g.\ \cite{good2020extreme, Fitkevich:2020okl,Myrzakul:2021bgj,Bianchi:2014qua}).  

We start with the entropy of a system in (1+1)-D CFT using $\epsilon$ as a UV cut-off \cite{Holzhey:1994we}, 
\begin{equation}
S=\frac{1}{6}\ln\frac{L}{\epsilon}\,, \label{conformal_entropy}   
\end{equation}
where $L$ is the size of the arbitrary system in general.  For us, the model will be the mirror trajectory which measures the size of the system by keeping track of accessible (1+1) dimensional spacetime in which the quantum field is free to propagate. That is, for a general and arbitrary moving mirror trajectory $p(u)$,
\begin{equation}
L\equiv p(u)-p(u_0)\,,  \label{general_size}  
\end{equation}
where $u$ and $u_0$ are null coordinates that form the region in the system which we are considering, and geometrically $\epsilon$ is asymmetrically smeared, i.e.\ $\epsilon^2\equiv\epsilon_p\epsilon_{p_0}$. 
Here $p(u)$ is the trajectory of the mirror in null coordinates (it is the usual advanced time function of retarded time $u$).  The smearing and dynamics of the mirror are related as 
\be
\epsilon_p = p'(u)\epsilon_u\,, \qquad
\epsilon_{p_0} = p'(u_0)\epsilon_{u_0}\,.\label{p'(u)} 
\ee
These $\epsilon_p$ smearings are coarse grained parameters at the end of the sub-systems, that parametrize
how well the observer distinguishes the subsystem from the rest of the universe \cite{Holzhey:1994we}.  

Substituting Eqs.~(\ref{p'(u)}) into Eq.~(\ref{conformal_entropy}) yields the bare entropy of the system,
\begin{equation}
S_{\textrm{bare}}=\frac{1}{12}\ln\frac{\left[p(u)-p(u_0)\right]^2}{p'(u)p'(u_0)\epsilon_u\epsilon_{u_0}}\ .    
\end{equation}
The vacuum entropy of the system can be found by considering a static mirror where 
$L=u-u_0$ and $\epsilon^2=\epsilon_u\epsilon_{u_0}$. Thus,
\begin{equation}
S_{\textrm{vac}}=\frac{1}{12}\ln\frac{(u-u_0)^2}{\epsilon_u\epsilon_{u_0}}\ .    
\end{equation}
Even though the entropies above are defined in terms of smearing, this dependence can be removed by an intuitive renormalization  (see also \cite{Bianchi:2014qua}) via 
\begin{equation}
S_{\textrm{ren}}\equiv S_{\textrm{bare}}-S_{\textrm{vac}}=\frac{1}{12}\ln\frac{[p(u)-p(u_0)]^2}{p'(u)p'(u_0)(u-u_0)^2}\ . \label{renorm_entropy}   
\end{equation} 
The geometric entropy Eq.~(\ref{renorm_entropy}) is invariant under M\"obius 
transformations, as seen from Eq.~\eqref{eq:jacob}. 
Thus a M\"obius transformation of mirror trajectories 
preserves energy flux, particle spectrum, and geometric 
entropy. 

Going further, if we define the derivative $p'(u_0)$ as 
\begin{equation} 
p'(u_0)=\frac{p(u)-p(u_0)}{u-u_0}\,, \label{expansion}  
\end{equation} 
then the dynamic meaning of the entropy becomes clear as 
we deviate $\delta u$ away from $u_0$.  We obtain the M\"obius invariant, 
\begin{equation}
S_{\textrm{ren}}=\frac{1}{12}\ln\frac{p'(u_0)}{p'(u)}=-\frac{1}{6}(\eta-\eta_0)\ .\label{renorm_entropy2}
\end{equation}
Choosing a static reference frame for time $u_0$ sets $p'(u_0)=1$ or $\eta_0 =0$, giving 
\begin{equation}
S(u)=-\frac{1}{12}\ln p'(u)=-\frac{\eta}{6}\ , \label{S(u)}
\end{equation} 
which is the common definition of entanglement 
entropy in terms of rapidity, e.g.\ \cite{Bianchi:2014qua,Chen:2017lum}. Interestingly Eq.~(\ref{S(u)}) is 
not preserved by a M\"obius transform. This is clear  
from $\eta(u)=(1/2)\ln p'(u)$ and $P(u)=(ap+b)/(cp+d)$, 
hence $P'=p'/(cp+d)^2$. The reason for this is that 
rapidity $\eta=\tanh^{-1}\dot z$ is not by itself invariant, but 
depends on the coordinate frame in which the mirror velocity 
$\dot z$ is measured. Both the invariant and variant von Neumann entropies, Eq.~(\ref{renorm_entropy2}) and Eq.~(\ref{S(u)}), measure the degree of quantum entanglement between the two subsystems, past and future. The more fundamental measure of entanglement of the system is Eq.~(\ref{renorm_entropy2}), which has invariance, like the particles and energy.  The operation of choosing a static mirror reference choice breaks the symmetry of the model: rapidity is always measured with respect to some frame.

\bibliography{main}

\end{document}